\documentclass{ws-procs9x6}

\begin{document}
\setcounter{page}{531}

\title{CONFORMATIONAL TRANSITIONS IN\\
MOLECULAR SYSTEMS\footnote{%
Work supported by Deutsche Forschungsgemeinschaft under grant No.~JA483/24-1/2.}}

\author{M.\ BACHMANN and W.\ JANKE}

\address{Institut f\"ur Theoretische Physik, Universit\"at Leipzig,\\
Postfach 100\,920, D-04009 Leipzig, Germany\\
E-mail: \{bachmann,janke\}@itp.uni-leipzig.de\\
www.physik.uni-leipzig.de/CQT.html}

\begin{abstract}
Proteins are the ``work horses'' in biological systems. In almost all functions 
specific proteins are involved. They control molecular transport processes,
stabilize the cell structure, enzymatically catalyze chemical reactions; others
act as molecular motors in the complex machinery of molecular synthetization 
processes. Due to their significance, misfolds and malfunctions of proteins 
typically entail disastrous diseases, such as Alzheimer's disease and
bovine spongiform encephalopathy (BSE). Therefore, the understanding of the trinity
of amino acid composition, geometric structure, and biological function is
one of the most essential challenges for the natural sciences. Here, we glance at
conformational transitions accompanying the structure formation in protein
folding processes.
\end{abstract}


\bodymatter

\section{Conformational Mechanics of Proteins}
\label{bj:sec1}
\begin{figure}[t]
\centerline{\epsfxsize=5cm \epsfbox{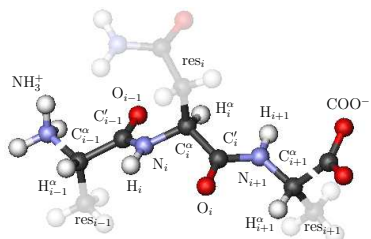}}
\caption{\label{bj:bb} 
Atomic composition of the protein backbone. Amino acids are
connected by the peptide bond between C$^\prime_{i-1}$ and N$_i$. Side chains or amino acid residues
(``res'') are usually connected to the backbone by a bond with the C$^\alpha_i$ atom
(except proline which has a second
covalent bond to its backbone nitrogen).}
\end{figure}

Structural changes of polymers and, in particular, proteins in collapse and crystallization
processes, but also in cluster formation and adsorption to substrates, require typically
collective and cooperative rearrangements of chain segments or monomers. Structure formation
is essential in biosystems as in many cases the function of a bioprotein is connected
with its three-dimensional shape (the so-called ``native fold''). 
Proteins are linear chains of amino acids linked by a peptide
bond (see Fig.~\ref{bj:bb}). Twenty different amino acids occur in biologically relevant, i.e., functional 
proteins. The amino acid residues differ in physical (e.g., electrostatic) and 
chemical (e.g., hydrophobic) properties. Hence, the sequence of amino acids typically entails 
a unique heterogeneity in geometric structure and, thus, a nonredundant biological function.

Proteins are synthesized by the ribosomes in the cell, where the genetic code in the DNA 
is translated into a sequence of amino acids. The folding of a synthesized protein into
its three-dimensional structure is frequently a spontaneous process. 
In a complex biological system, the large variety of processes which are necessary to 
keep an organism alive requires a large number of different functional proteins. In the 
human body, for example, about 100\,000 different proteins fulfil specific functions.
However, this number is extremely small, compared to the huge number of possible 
amino acid sequences ($=20^N$, where $N$ is the chain length and is typically between 100 and 3000).
The reason is that bioproteins have to obey very specific requirements. Most important are
stability, uniqueness, and functionality. 

\begin{figure}[t]
\centerline{\epsfxsize=5cm \epsfbox{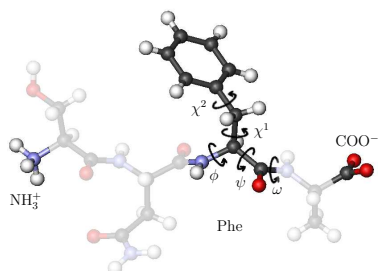}}
\caption{\label{bj:struct} 
Definition of the backbone dihedral angles $\phi$, $\psi$, and $\omega$. Exemplified for
phenylalanine, also the only two side-chain degrees of freedom $\chi^1$ and $\chi^2$ are
denoted. The convention is that the torsional angles can have values between $-180^\circ$
and $+180^\circ$, counted from the N-terminus (NH$_3^+$)
to the C-terminus (COO$^-$) according to the right-hand rule
and in the side chains starting from the C$^\alpha$ atom.
}
\end{figure}

Under physiological conditions, flexible protein degrees
of freedom are the dihedral angles, i.e., a subset of backbone and 
side-chain torsional angles (see Fig.~\ref{bj:struct}). 
Denoting the set of dihedral angles of the $n$th amino acid in the chain by
$\boldsymbol{\xi}_n=\{\phi_n,\psi_n,\omega_n,\chi^{(1)}_n,\chi^{(2)}_n,\ldots \}$,
the conformation of an $N$ residue protein is then entirely defined by 
${\bf X} = {\bf X}(\boldsymbol{\xi}_1, \boldsymbol{\xi}_2,\ldots,\boldsymbol{\xi}_N)$.
Therefore, the partition function can formally be written as a path integral over
all possible conformations:
\begin{equation}
\label{bjeq:pf}
Z=\int {\cal D}{\bf X}\exp\left[-E({\bf X})/k_BT\right],\quad \int {\cal D}{\bf X}=
\prod\limits_{n=1}^N\left[\int d\boldsymbol{\xi}_n\right],
\end{equation}
where $E(\bf X)$ is the energy of the conformation $\bf X$ in
a typically semiclassical all-atom protein model. A precise modeling is intricate because of the 
importance of quantum effects in this complex macromolecular system, which are ``hidden'' in the
parametrization of the semiclassical model. Another important problem is the modeling of
the surrounding, strongly polar solvent. The hydrophobic effect that causes the formation of
a compact core of hydrophobic amino acids screened from the polar solvent by a shell of
polar residues is expected to be the principal driving force towards the native, functional
protein conformation.~\cite{dill1,still1,bj1} Conformational transitions accompanying molecular 
structuring processes, however, exhibit similarities to thermodynamic phase transitions
and it should thus be possible to characterize these transitions by means of a 
strongly reduced set of effective degrees of freedom, in close correspondence to 
order parameters that separate thermodynamic phases. 
Assuming that a single ``order'' parameter $Q$ is sufficient to distinguish between two (pseudo)phases,
its mean value 
$\langle Q\rangle=Z^{-1}\int {\cal D}{\bf X}\, Q({\bf X})\exp\left[-E({\bf X})/k_BT\right]$
should possess significantly different values in these phases. In typical
first-order-like nucleation transitions such as helix formation~\cite{gbcj1} 
or tertiary two-state folding~\cite{kbj1}, the free-energy landscape 
$F(Q)\sim-k_BT\ln\left\langle \delta(Q-Q_0({\bf X}))\right\rangle$ exhibits 
a single folding barrier.
\section{From Microscopic to Mesoscopic Modeling}
\label{bj:sec2}
If the characterization of conformational macrostates by low-dimensional parameter spaces is
possible, it should also be apparent to introduce coarse-grained substructures and thus to 
reduce the complexity of the model to a minimum. Such minimal models for proteins have indeed been
introduced~\cite{dill1,still1} and have proven useful in 
thermodynamic analyses of folding, adsorption, and 
aggregation of polymers and proteins.~\cite{bj1,ssbj1,bj2,jbj1,kbj1} 

In the simplest approaches~\cite{dill1,still1}, only
two types of amino acids are considered: hydrophobic and polar residues. This is plausible as 
most of the 20 amino acids occurring in natural bioproteins can be classified with respect
to their hydrophobicity. Amino acids with charged side chains or with residues containing polar 
groups (amide or hydroxylic) are soluble in the aqueous environment, because these groups are 
capable of forming hydrogen bonds with water molecules. Nonpolar amino acids
do not form hydrogen bonds and, if exposed to water, would disturb the hydrogen-bond network. This is
energetically unfavorable. In fact, hydrophobic amino acids effectively attract each other and
typically form a compact hydrophobic core in the interior of the protein. 
\begin{figure}[t]
\vspace*{-0.2cm}
\centerline{\epsfxsize=4.7cm \epsfbox{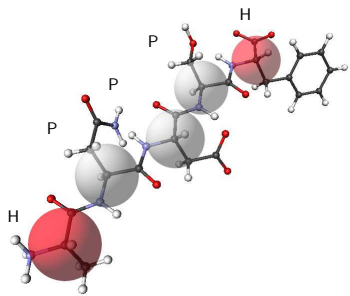}}
\caption{\label{bj:meso} 
Coarse-graining proteins in a ``united atom'' approach. Each amino acid is
contracted to a single ``C$^\alpha$'' interaction point. The effective distance
between adjacent, bonded interaction sites is about 3.8\,\AA. 
In the class of so-called hydrophobic-polar models, only hydrophobic (H)
and polar (P) amino acid residues are distinguished.}
\end{figure}

Figure~\ref{bj:meso} shows an example how the complexity of a protein segment
can be reduced by coarse-graining. On one hand, the 
residual complexity is limited by only distinguishing hydrophobic (H) and 
polar (P) amino acids. On the other hand, the steric extension of the side chains is
mesoscopically rescaled and the whole side chain is contracted into a single interaction point.
Volume exclusion in the interaction of different side chains is then energetically modeled
by short-range repulsion. For this reason, lattice proteins are modeled as self-avoiding
walks~\cite{dill1} and in off-lattice models Lennard-Jones-like potentials~\cite{still1} 
satisfy this constraint. 

Systematic enumeration studies 
of simplified hydrophobic-polar lattice models have indeed qualitatively 
revealed characteristic features of real proteins, such as the small number of amino acid sequences
possessing a unique native fold, but also the comparatively small number of native topologies
proteins fold into.~\cite{sbj1} It is also remarkable that typical protein folding paths
known from nature are also identified by employing coarse-grained models. This regards, in particular,
folding landscapes with characteristic barriers~-- from the simple two-state characteristics
with a single kinetic barrier,~\cite{kbj1} over folding across several barriers via weakly stable
intermediate structures, to folding into degenerate native states.~\cite{ssbj1} Metastable 
conformations as in the latter case are important for biological functions, where the 
local refolding of protein segments is essential, as, e.g., in molecular motors.
\section{A Particularly Simple Example: Two-State Folding}
\begin{figure}[t]
\centerline{\epsfxsize=7cm \epsfbox{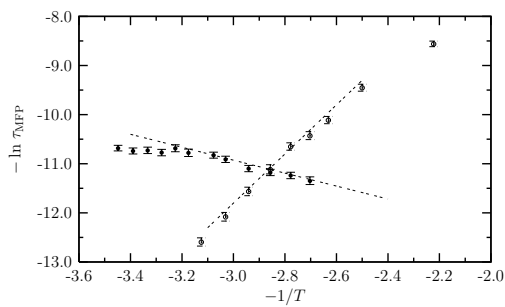}}
\caption{\label{bj:chev}
Chevron plot of the mean-first passage times from
folding ($\bullet$) and unfolding ($\circ$) events at different temperatures. 
The hypothetic intersection point
corresponds to the transition state.~\cite{kbj1}
}
\end{figure}

A few years ago, experimental evidence was found that classes of proteins show
particular simple folding characteristics, single exponential and two-state
folding~\cite{fersht1}.
In the two-state folding process, the peptide is
either in an unfolded, denatured state or it possesses a native-like, folded structure.
In contrast to the barrier-free single-exponential folding, there exists an unstable
transition state to be passed in the two-state folding process. This can nicely be seen
in the exemplified chevron plot shown in Fig.~\ref{bj:chev}, obtained from Monte Carlo computer
simulations of folding and unfolding events of a mesoscopic protein model.~\cite{kbj1}
In this plot, the mean first passage (MFP) time $\tau_\text{MFP}$ (in Monte Carlo steps)
is plotted versus temperature. The MFP time is obtained by averaging the times passed in the 
folding process from a random conformation to the stable fold over many folding
trajectories. MFT times for unfolding events can be estimated in a like manner, but one starts from
the native conformation and waits until the protein has unfolded. A structure is defined
to be folded, if it is structurally close to the native conformation. A frequently used measure
is the fraction $Q$ of already established native contacts (i.e., the number of residue pairs 
that reside within the optimal 
van der Waals distance), 
compared to the total number of contacts
the native fold possesses. Thus, if $Q>0.5$, the structure is folded and unfolded if $Q<0.5$. For
$Q=0.5$, the conformation is in the transition state. Apparently, $Q$ serves as a sort of order parameter.

The two branches in Fig.~\ref{bj:chev}
belong to the folding and 
unfolding events. With increasing temperature
folding times grow, and unfolding is getting slower with decreasing temperature. These two processes
are in competition with each other and the intersection point defines the folding transition temperature.
The whole process exhibits characteristics of first-order-like phase transitions. At the intersection
point, the ensembles of folded and unfolded conformations coexist with equal weight. 
In the transition region, both branches exhibit exponential behavior. 
Thus, $\tau_\text{MFP}$ is directly related 
to exponential folding and unfolding rates
$k_{f,u}\approx 1/\tau_\text{MFP}^{f,u}\sim \exp(-\varepsilon_{f,u}/k_BT)$, respectively, where the constants
$\varepsilon_{f,u}$ determine the kinetic folding (unfolding) propensities. The dashed lines in
Fig.~\ref{bj:chev} are tangents to the logarithmic folding and unfolding curves at the transition state
temperature.
\section{Conclusion}
Conformational transitions of macromolecular systems, in particular, proteins, exhibit clear
analogies to phase transitions in thermodynamics. The main difference is that proteins
are finite systems and a thermodynamic limit does not exist. Nonetheless, the analysis
of structure formation processes in terms of an ``order'' parameter is also 
a very useful approach to a better understanding of conformational transitions.
In this context it also turns out to be reasonable to introduce coarse-grained models
where the reduction to only relevant degrees of freedom allows for a more systematic 
analysis of characteristic features of 
protein folding processes than it is typically possible with models containing specific 
properties of all atoms.
%
%
%

%
\end{document}